# Raman Spectroscopy of Graphene under Uniaxial Stress: Phonon Softening and Determination of the Crystallographic Orientation

Mingyuan Huang[1*], Hugen Yan[2*], Changyao Chen[1], Daohua Song[2], Tony F. Heinz[2], and James Hone[1†]

[1]*Department of Mechanical Engineering, Columbia University, New York, New York, 10027 USA*

[2]*Departments of Physics and Electrical Engineering, Columbia University, New York, New York, 10027 USA*

*These authors contributed equally to this work

[†]e-mail:jh2228@columbia.edu

**We present a systematic study of the Raman spectra of optical phonons in graphene monolayers under tunable uniaxial tensile stress. Both the G and 2D bands exhibit significant red shifts. The G band splits into two distinct sub-bands ($G^+$, $G^-$) because of the strain-induced symmetry breaking. Raman scattering from the $G^+$ and $G^-$ bands shows a distinctive polarization dependence that reflects the angle between the axis of the stress and the underlying graphene crystal axes. Polarized Raman spectroscopy therefore constitutes a purely optical method for the determination of the crystallographic orientation of graphene.**

Since the discovery of mechanical cleavage of graphene from graphite crystals(1), graphene has attracted intense interest because of properties that include high electron mobility(2, 3), novel quantum Hall physics(4, 5), superior thermal conductivity(6), and unusually high mechanical strength(7).  Raman spectroscopy has emerged as a key diagnostic tool to identify single layer graphene sheets(8) and probe their physical properties(9, 10). Because strain induces shifts in the vibrational frequencies, Raman spectroscopy can be applied to map built-in strain fields during synthesis(11) and device fabrication, as well as measure load transfer in composites.  The rate of shift of the phonon frequencies with strain depends on the anharmonicity of the interatomic potentials of the atoms in the honey-comb lattice and thus can be used to verify theoretical models.

Measurement of the strain dependence of the Raman active phonons is thus important for both applied and fundamental studies of this material system.  Using graphene supported on a flexible substrate, we have been able to obtain precise information on the rate of frequency shift of the Raman G (zone-center optical) and 2D (two-phonon zone-edge optical) modes with strain. In addition, the polarization dependence of the Raman response in strained graphene can, as we demonstrate in this paper, be used for an accurate determination of the crystallographic orientation.  For unstrained graphene, such an orientation analysis is precluded by the high symmetry of the hexagonal lattice. A particularly important application of this capability lies in the study of nanopatterned graphene monolayers, such as nanoribbons(12) and quantum dots(13). Graphene nanoribbons possess electronic band gaps whose magnitude reflects both the ribbon width and crystallographic orientation(12, 14-16). The electronic states associated with graphene edges are also sensitive to the crystallographic orientation of the ribbon(17). It is thus crucial to be able to correlate the measured properties to the underlying crystallographic orientation of the sample. As we show here, polarized Raman spectroscopy provides a simple, but precise analytic tool that complements electron-spectroscopy techniques such as scanning tunneling microscopy (STM)(18), transmission electron microscopy (TEM)(19), and low-energy electron diffraction (LEED)(20), methods that typically require ultrahigh vacuum conditions and specialized equipment.

Figures 1a and 1b show the 2D and G bands of graphene under increasing strain. As expected, both bands exhibit significant red shifts. Further, the G mode peak splits as the graphene symmetry is lowered by the strain. Unstrained graphene belongs to the two-dimensional (2-D) point group $D_6$ and displays $C_6$ and $m$ symmetry. Under strain only $C_2$ symmetry will generally be retained. (If the strain is in the armchair or zigzag direction, $m$ symmetry will also survive.) Therefore, the G band, which results from a doubly-degenerate optical phonon mode with $E_{2g}$ symmetry, splits into two singlet bands, which we denote by $G^+$ and $G^-$, according to their energy. Fig. 1c shows the positions of the 2D, $G^+$, and $G^-$ peaks versus applied strain. All three modes shift linearly with strain, with shift rates of 21, 12.5, and 5.6 cm$^{-1}$/% for the 2D, G$^-$, and G$^+$ bands, respectively. In addition, we find that the strain-induced symmetry breaking reveals previously unexamined physics of the G mode. Therefore, below we focus exclusively on the G mode.

The G band of graphene is a long wavelength (q = 0) optical phonon mode in which the two sub-lattices vibrate with respect to each other. Therefore, a phenomenological method can be used to analyze the Raman modes in the presence of strain(21). Following the standard procedure (see supporting information), we can write the secular equation for the G band of strained graphene as

$$\begin{vmatrix} A\varepsilon_{xx} + B\varepsilon_{yy} - \lambda & (A-B)\varepsilon_{xy} \\ (A-B)\varepsilon_{xy} & B\varepsilon_{xx} + A\varepsilon_{yy} - \lambda \end{vmatrix} = 0 \qquad (1)$$

where $\lambda \equiv \omega^2 - \omega_0^2$ is the difference between the square of the perturbed and unperturbed phonon frequencies, $A$ and $B$ are the phonon deformation potential coefficients, and $\varepsilon_{ij}$ is the strain tensor. Because second- and fourth-order tensors describing the in-plane response of the hexagonally symmetric graphene have the same form as those of isotropic materials, the secular equation holds in any coordinate system in the graphene plane. In the experimental coordinate system, the strain tensor can be reduced to $\varepsilon_{xx} = \varepsilon$, $\varepsilon_{yy} = -\upsilon\varepsilon$, and $\varepsilon_{xy} = 0$, where $x$ is the direction of the applied stress and $\varepsilon$ is the magnitude of the resulting strain and $\upsilon = 0.16$ is the Poisson ratio(22). We then find that the frequencies of the $G^+$ and $G^-$ modes are given by

$$\begin{cases} \omega_{G^+} = \omega_0 + \dfrac{B - \upsilon A}{2\omega_0}\varepsilon \\ \omega_{G^-} = \omega_0 + \dfrac{A - \upsilon B}{2\omega_0}\varepsilon \end{cases} \quad (2)$$

The corresponding eigenvectors for the atomic displacements are perpendicular (transverse) to the direction of stress for the $G^+$ mode, and parallel (longitudinal) for the $G^-$ mode.

Using the measured shift rates, we find for the phonon deformation potential coefficients $A = -4.4 \times 10^6$ cm$^{-2}$ and $B = -2.5 \times 10^6$ cm$^{-2}$ of the G band. To compare the results with previous experiments on graphite performed under three-dimensional hydrostatic pressure, we use the values of A and B to calculate the shift rate of graphene under two-dimensional hydrostatic pressure. The inferred shift rate is 1.9 cm$^{-1}$/GPa, smaller than the value measured from graphite (4.6 cm$^{-1}$/GPa)(23). For graphene, the resulting G-mode Grüneisen constant, which measures its anharmonicity, is 0.67. This value is smaller than the corresponding Grüneisen constant for graphite, which is 1.59 after normalization to two dimensions(23, 24). The larger value for graphite may reflect its interlayer interactions, which are strongly altered by pressure.

Polarized Raman spectroscopy can give insight into crystal orientation and vibrational symmetry. We have studied the polarization of the $G^+$ and $G^-$ bands with the incident light polarized parallel to the principal strain axis. Figures 2a and 2b present two-dimensional plots of the Raman intensity as a function of Raman shift and angle $\phi$ between the incident and scattered light polarization for two different graphene samples. Both the $G^+$ and $G^-$ bands exhibit strong polarization dependence. This stands in contrast to the isotropic Raman response of the G band for unstrained graphene. For sample 1 (Fig. 2a), the $G^+$ band intensity peaks near $\phi = 75°$ and nearly vanishes for $\phi = 165°$. The polarization of the scattered Raman radiation in the $G^-$ band is simply shifted by 90°. The results of sample 2 (Fig. 2b) are similar, but with the maxima occurring at different angles. These findings indicate that the scattered light polarization is affected by the crystal orientation. To quantify the polarization properties of the Raman scattering process for $G^+$ and $G^-$ bands, their intensity is plotted in Figs 2c and 2d as a function of the angle of the analyzer. The polarization dependence can be fit by the form $\sin^2(\phi - \phi_0)$ with $\phi_0 =$ 165.3°, 75.8° for the $G^+$ and $G^-$ modes, respectively, for sample 1, and with $\phi_0 = 53.6°$, 140° for sample 2. This indicates that the scattered light from $G^+$ and $G^-$ bands is linearly polarized, and that the two modes have orthogonal polarizations.

According to the usual semi-classical treatment, first-order Raman scattering arises from the derivative of the electric susceptibility with respect to the atomic displacement of the relevant vibration. The polarization dependence of the scattering intensity can then be expressed as $I_s \propto |\mathbf{e}_i R \mathbf{e}_s|^2$ (25), where $\mathbf{e}_i$ and $\mathbf{e}_s$ are the unit vectors describing the polarizations of the incident and scattered light. The Raman tensor $\mathbf{R}$ is determined by the symmetries of crystal and vibrational mode. The G mode of unstrained graphene is doubly degenerate. Two Raman tensors (in the crystal reference system, Fig. 2f) contribute to the total scattering intensity:

$$R_x = \begin{bmatrix} 0 & d \\ d & 0 \end{bmatrix} \quad R_y = \begin{bmatrix} d & 0 \\ 0 & -d \end{bmatrix} \quad (3)$$

Here $R_x$ and $R_y$ correspond to the modes in which carbon atoms vibrate along the lattice $x$ and $y$ directions. These modes scatter light in such a way that the polarization vector is reflected, respectively, about the lines $x = y$ and $y = 0$. Because the scattering intensities are equal for the two modes, this process completely depolarizes the inelastically scattered light, rendering the G band response independent of polarization.

As discussed above, strain splits the G band into $G^-$ and $G^+$ bands whose normal modes are parallel and perpendicular to the strain axis, respectively. The polarization of the scattered light from each mode will then depend on the direction of the strain relative to the crystal lattice. For example, we consider two cases in which the incident light is polarized parallel to the strain. If the strain is in the 'zigzag' direction (Fig. 2e), the $G^-$ mode is equivalent to $R_x$, and the scattered light is polarized perpendicular to the strain direction. If the strain is in the 'armchair' direction (Fig. 2f), the $G^-$ mode is equivalent to $R_y$, and the scattered light is polarized parallel to the strain direction. Conversely, the $G^+$ band corresponds to $R_y$ for strain along the zigzag axis and to $R_x$ for strain along the armchair axis, leading to polarization perpendicular to that of the $G^-$ mode.

For quantitative analysis of the polarization, we consider a general case in which the strain is at an arbitrary angle $\theta$ with respect to the graphene crystallographic *x*-axis. The Raman tensors of the $G^+$ and $G^-$ bands are given by linear combinations: $R^\pm = v_x^\pm R_x + v_y^\pm R_y$, where ($v_x^+$, $v_y^+$), ($v_x^-$, $v_y^-$) are the unit vectors along the vibration direction of the $G^+$ and $G^-$ bands respectively. If the polarization vectors $\mathbf{e}_i$, $\mathbf{e}_s$ are assumed to make arbitrary angles $\psi$, $\phi$ with respect to strain axis, the intensities of the $G^+$ and $G^-$ bands will then vary as (see supporting information for details)

$$\begin{cases} I_{G^+} \propto d^2 \sin^2(\phi + \psi + 3\theta) \\ I_{G^-} \propto d^2 \cos^2(\phi + \psi + 3\theta) \end{cases} \quad (4)$$

This result is consistent with the experimental observation that the $G^+$ and $G^-$ bands produce linearly polarized radiation with orthogonal polarization states. In addition, Eq. (4) shows that the polarization of the Raman scattered light is determined by the angle between the strain axis and the graphene crystal axes (with $\psi = 0$ for the experimental conditions). Analysis of the polarization dependence of the Raman scattering process thus provides an optical method of determining the orientation of the graphene crystallographic axes. Comparing our experimental results to Eq. (4), we conclude that for sample 1 the strain is applied at an angle of 25.2±0.1° from the zigzag direction, while for sample 2 the strain is applied at 12.7±0.6° from the zigzag direction.

To further confirm the theoretical analysis, an additional measurement was performed. We rotated sample 1 about its surface normal, while the incident and detected light polarizations remained fixed and perpendicular to one another. The data (Fig. 3a) show that the period of variation in the scattering intensity from each band is 90º. This result is consistent with Eq. (4), since rotation of the sample is equivalent to a simultaneous and equal change of $\psi$ and $\phi$. The maximal intensities of the $G^+$ and $G^-$ bands (Fig. 3b) are equal to one another, and about half of the G band intensity measured without a polarizer. This behavior is consistent with the Raman tensor analysis above and indicates that the strain does not significantly change the Raman scattering intensity.

Finally, we note that both the $G^+$ and $G^-$ bands have the same spectral width as that of the G band without applied strain. This finding implies that the strain within the laser spot (~1 micron) is very homogeneous, consistent with the results of finite element simulation (see supporting information). It also indicates that the electron-phonon coupling strength, which largely defines the linewidth, is equivalent for $G^+$ and $G^-$ phonons. The last result stands in contrast to the case of metallic carbon nanotubes, whose $G^+$ and $G^-$ modes exhibit distinct coupling to electron-hole pair excitations(10, 26). This difference arises from the one-dimensional structure of nanotubes, which acts to enhance the LO phonon interaction with electrons(27).

An important potential application of this technique involves the fabrication of graphene nanoribbons and similar devices that are oriented along a pre-determined crystal axis. For these applications, one would like to determine the lattice orientation directly on a Si wafer without transferring the graphene sample to a flexible substrate. From Fig. 1b, we observe that at least 1.5% strain is needed to separate $G^+$ and $G^-$ bands fully from one another. However, such a complete separation of the peaks is not necessary for determining the crystal orientation: For smaller strains, the apparent position of the G band will vary with the detected Raman polarization. Assuming that a modulation of ~2 cm$^{-1}$ can be observed, the crystal orientation could then be measured with an applied strain of about 0.3%, which can be achieved on Si substrates(28).

In summary, we have transferred monolayer graphene onto flexible substrates and measured the Raman spectrum under uniaxial tensile strain. The 2D and G bands exhibit red shifts and the G band splits into two distinct ($G^+$, $G^-$) features because of the strain-induced breaking of the crystal symmetry. Unlike for the G mode of unstrained graphene, Raman scattering from the $G^+$ and $G^-$ modes exhibits strong polarization dependence, with a response that depends on the orientation of the graphene crystal lattice. This behavior permits a precise determination of the crystallographic orientation of graphene monolayers by means of a purely optical technique.

**Materials and Methods**

The graphene samples investigated in this study were prepared by mechanical exfoliation of Kish graphite(1). Monolayer flakes, deposited on Si wafers with a 290 nm $SiO_2$ epilayer, were initially identified by optical microscopy and their nature was confirmed by Raman spectroscopy(8). After a suitable graphene flake was selected, we transferred the graphene monolayer from the Si wafer onto the surface of a polydimethylsiloxane (PDMS) film, a clear silicone elastomer suitable for the application of stress. The transfer technique, outlined in Figs 4a-4d, is similar to that previously demonstrated for carbon nanotubes(29). Figures 4e and 4f show optical images of a graphene flake on the $SiO_2$ surface (before transfer) and on the PDMS film (after transfer). We note that this method can be used to transfer graphene from the $Si/SiO_2$ substrate, where it is particularly easy to identify, to any other substrate. The graphene monolayer is still visible on the transparent PDMS substrate, a result that can be confirmed by a reflectance calculation(30). In order to clamp the graphene onto the PDMS film after its transfer, we patterned narrow strips of titanium (60 nm thick, 2 μm wide) on the samples by evaporation through a shadow mask (Fig. 4g).

Uniaxial stress was applied to the graphene in a direction perpendicular to the Ti strips by controlled bending of the PDMS. The induced longitudinal strain was determined directly by measuring the distance between the strips in an optical microscope. For the transverse direction, the strain was calculated from the Poisson ratio (see supporting information). The applied strain was completely reversible, indicating that there was no slippage or permanent modification of the sample when stressed. For Raman spectroscopy, a 532-nm excitation laser was focused on the graphene monolayer by a 40× microscope objective with a NA of 0.52. The Raman shifted light was collected by the same objective in a backscattering configuration. The laser power was kept sufficiently low (<1 mW) to avoid heating during the measurement.

**Acknowledgments**

We acknowledge support from the Nanoscale Science and Engineering Initiative of the NSF (NIRT grant ECS-05-07111), the DARPA Center on Nanoscale Science and Technology for Integrated Micro/Nano-Electromechanical Transducers (iMINT, grant HR0011-06-1-0048).

**Competing financial interests**

The authors declare that they have no competing financial interests.

**Fig. 1.** Strain induced phonon softening in graphene. a, b, Evolution of the spectra of the 2D (a) and G (b) bands of graphene under strain. The spectra of the 2D band and the first two spectra of the G band are fit by single Lorentz peaks; the other G-band spectra are fit by two Lorentz peaks of fixed width 16 cm$^{-1}$ (smooth overlapping curves), as extracted from the following polarization study (Fig. 3b). The difference between the G band data and the fits is attributed to scattering from the PDMS film. c, The variation of the phonon frequencies of 2D, G$^+$ and G$^-$ bands from a and b as a function of strain. The solid lines are linear fits.

**Fig. 2.** Polarization analysis of the G band of strained graphene. a, b, False-color image of the intensity of the Raman scattered light as a function of the Raman shift and angle of the analyzer $\phi$ for detection of the Raman signal. The angle $\phi$ was measured with respect to the incident light polarization, which was aligned along the strain axis. The data were obtained by measuring Raman spectra every 10° for sample 1 ($\varepsilon = 2.3\%$, a) and 2 ($\varepsilon = 1.9\%$, b). c, d, Raman scattering intensity for the G$^+$ (blue triangle) and G$^-$ (red triangle) bands as a function of $\phi$ for samples 1 and 2, respectively. The solid lines are fits to the form of $\sin^2(\phi-\phi_0)$. e, f, Schematic representation of the vibrational modes for the G$^-$ bands when the strain axis (green arrow) is, respectively, in the zigzag (e) and the armchair (f) directions of the graphene crystal. The yellow arrows represent the polarization of the incident light, which is chosen as parallel to the strain axis. The red arrows show the resulting polarization of the Raman scattered light for the G⁻ mode. The coordinate system in f represents the crystal reference system.

**Fig. 3.** The dependence of Raman spectra on the sample orientation. a, False color image of the Raman intensity as a function of the Raman shift and the angle $\phi$ of rotation of the strained graphene sample ($\varepsilon = 2.3\%$) about its surface normal. The magenta curve, which acts as a guide to the eye, is the intensity of the G⁻ or G⁺ bands (measured to the left and right, respectively). The polarizations of the pump radiation and detected Raman radiation were fixed and perpendicular to one another. b, Comparison of the spectra of the pure G⁻, G⁺ and G bands. The spectra of the G⁻ (black circles, obtained for $\phi = 50°$ ) and G⁺ (red squares, obtained for $\phi = 0°$) are included in (a), while the spectrum of the G mode (blue triangles) was measured on the same sample without strain. The fits (solid lines) are Lorentzian in shape.

**Fig. 4.** Graphene transfer process. a-d, Schematic illustration of the transfer of graphene from Si wafer to the PDMS film. The initial step involved deposition of a gold film (yellow) on the Si wafer supporting graphene monolayer; concentrated polyvinyl alcohol (PVA) solution was then cast onto the gold film (a). After the PVA solidified, it was peeled off from the Si wafer, carrying with it the gold and graphene films (b), which were then transferred onto PDMS substrate (c). In the last step, the PVA was washed away by DI water and the gold was dissolved by etching (d). e-g, The same graphene on the Si wafer (before transfer) (e), on the PDMS substrate (after transfer) (f), and as clamped by Ti strips (g).

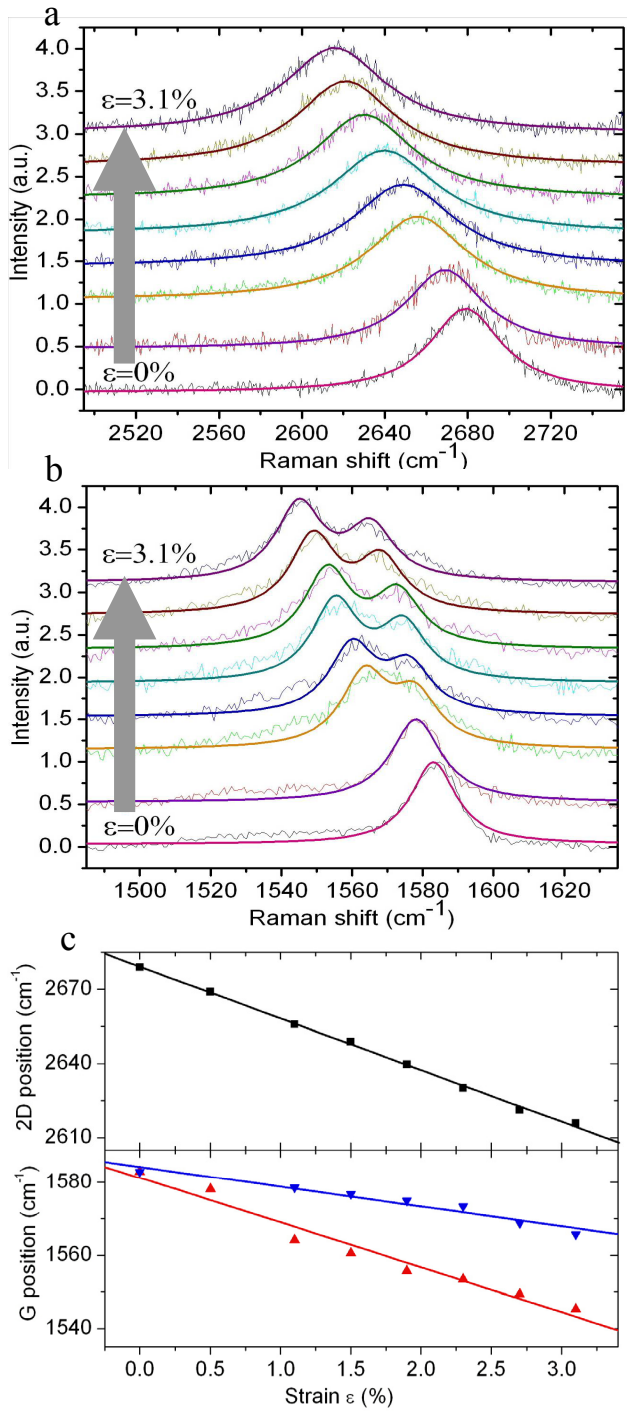

Huang et al. Figure 2

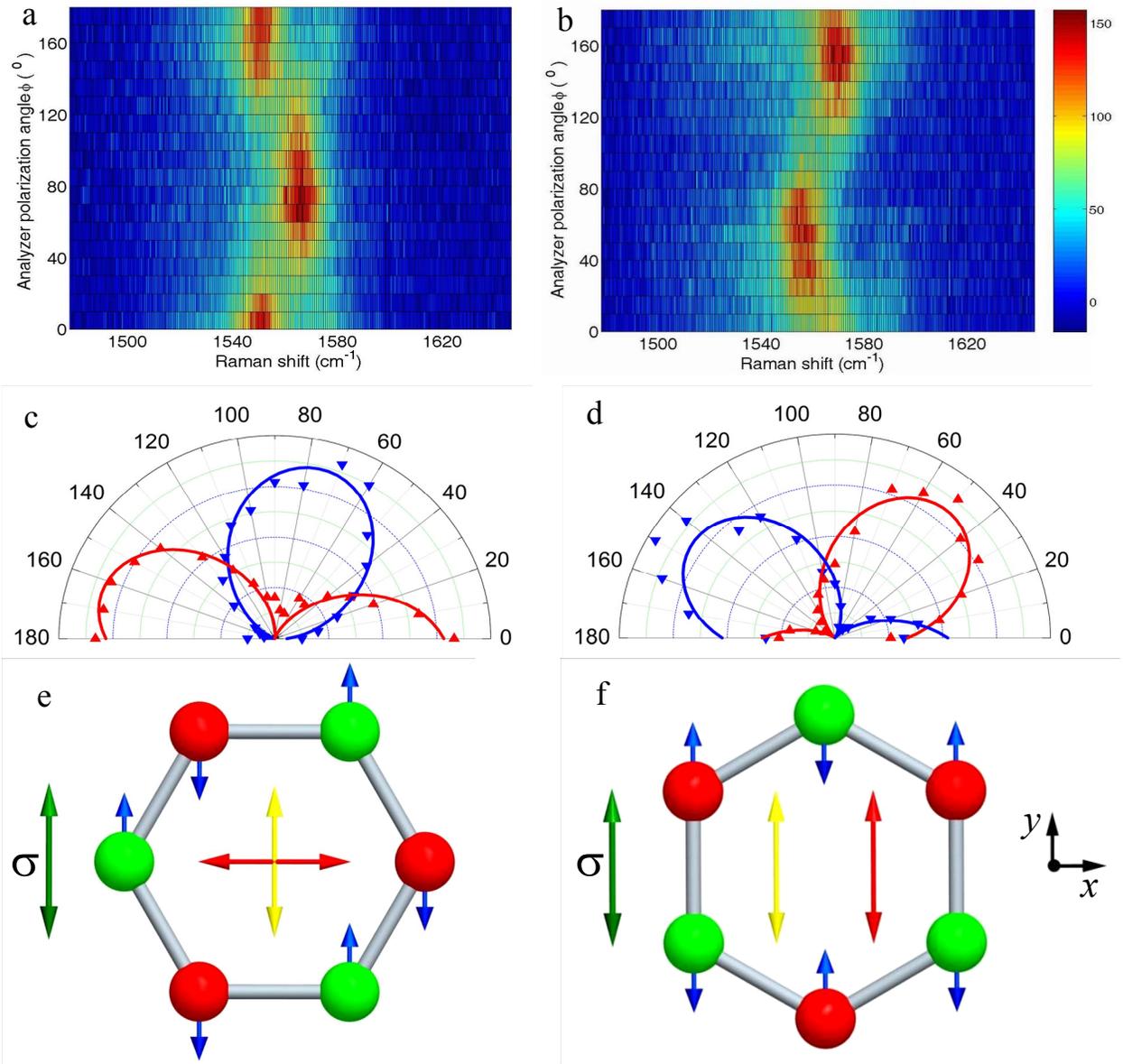

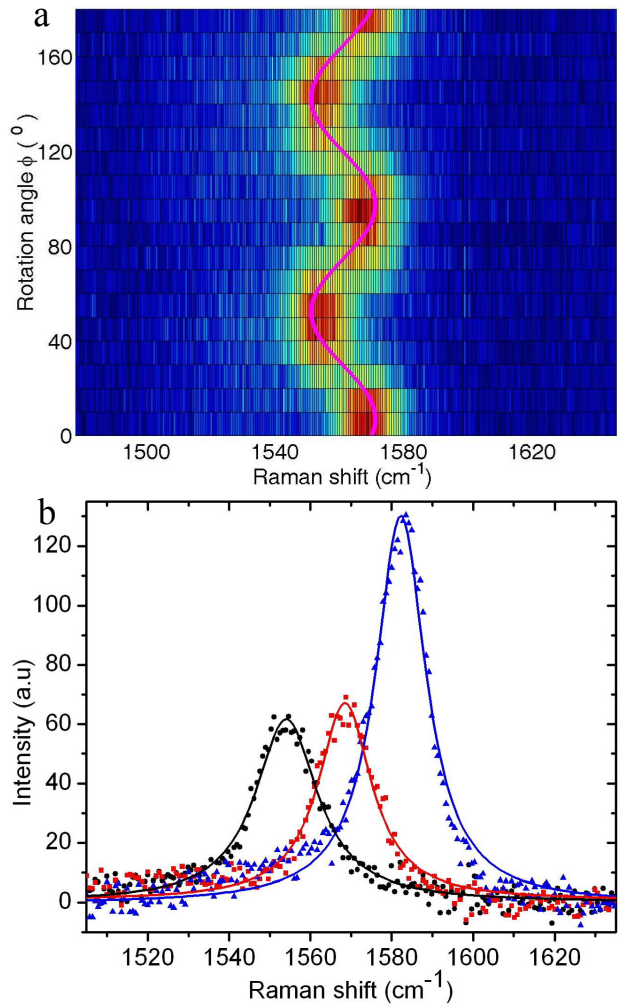

Huang et al. Figure 4

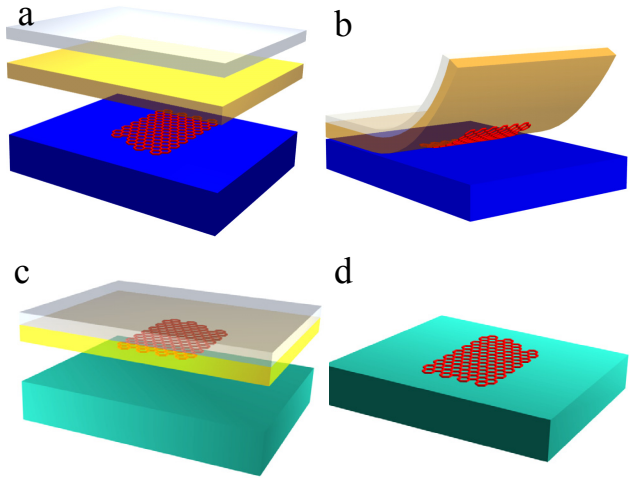
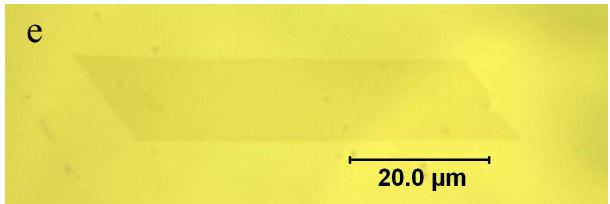
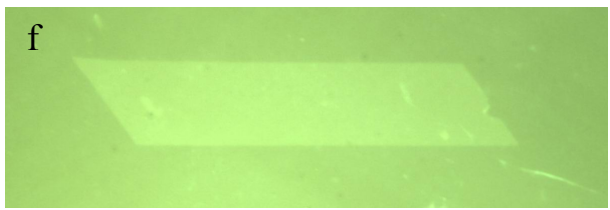
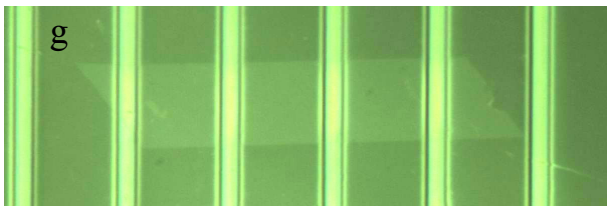